\documentclass[cameraready]{Interspeech}

\title{DNSMOS-C: Improving End-to-end Speech Quality Models via Contrastive Learning}

\author[affiliation={1}]{Xinyu}{Liang}
\author[affiliation={1}]{Fredrik}{Cumlin}
\author[affiliation={2}]{Victor}{Ungureanu}
\author[affiliation={2}]{Chandan K. A.}{Reddy}
\author[affiliation={2}]{Christian}{Sch\"uldt}
\author[affiliation={1}]{Saikat}{Chatterjee}

\address{
    $^1$ KTH Royal Institute of Technology, Stockholm, Sweden \\
    $^2$ Google LLC
}

\email{hopeliang@icloud.com, fcumlin@kth.se, ungureanu@google.com, chandanka@google.com, cschuldt@google.com, sach@kth.se}


\usepackage{comment}
\usepackage{adjustbox}
\usepackage{subcaption}
\usepackage{multirow}

\begin{document}

\maketitle

\begin{abstract}
    We introduce DNSMOS-C, a compact end-to-end speech quality assessment model that extends the DNSMOS Pro framework by integrating a MOS-guided triplet-based contrastive loss. Applied directly to the intermediate embeddings, this contrastive supervision encourages the latent space to be better organized with respect to perceptual quality while preserving the simplicity and efficiency of DNSMOS Pro. Unlike prior methods that depend on large pre-trained self-supervised learning (SSL) encoders and multi-stage training, DNSMOS-C jointly learns speech representations and MOS regression within a single, unified framework. Experiments on multiple datasets show that DNSMOS-C consistently improves correlation metrics over DNSMOS Pro and achieves better generalization on challenging out-of-domain test sets. Furthermore, latent space analyses indicate that our approach learns representations that exhibit an emergent low-dimensional quality ordering, which enhances interpretability and improves training stability. These findings demonstrate that MOS-guided contrastive learning enables more robust and accurate quality predictions without incurring additional computational overhead.
\end{abstract}

\section{Introduction}
\label{sec:intro}

Accurate and automatic speech quality assessment (SQA) plays a critical role in developing and monitoring modern audio technologies, ranging from streaming services to generative speech models. Traditionally, subjective evaluations such as mean opinion scores (MOS) provide the gold standard for quality assessment, but they are time-consuming, expensive, and not scalable. Consequently, recent research has focused on developing objective non-intrusive SQA models that predict MOS directly from degraded speech without requiring clean reference signals.

Large-scale models leveraging self-supervised learning (SSL) representations \cite{SSL-MOS, UTMOS, SSL-Layer-MOS, multivariate} or multimodal large language models (LLMs) \cite{wang2025enabling} have achieved state-of-the-art SQA performance. However, these approaches come with significant computational and memory overheads, making them less suitable for real-time deployment or resource-constrained environments.

To address efficiency concerns, compact end-to-end SQA models based on convolutional architectures have become the standard~\cite{MOSNet, DeePMOS, NISQA, LDNet}. Among these, DNSMOS \cite{DNSMOS} and its successor, DNSMOS Pro \cite{DNSMOSp}, demonstrate superior performance and efficiency. This is largely achieved by replacing the recurrent network module with a global pooling layer, making them highly suitable for real-time applications such as VoIP deployment. However, a significant limitation of these models is their still limited ability to generalize to unseen distortions or recording conditions \cite{generalization}.

Recent advances in contrastive learning have demonstrated strong potential for learning well-structured and semantically meaningful embeddings \cite{hermans2017defense, chen2022learning, wisnu2025improving}. In speech quality assessment, SCOREQ \cite{SCOREQ} introduced a triplet-based contrastive regression loss that organizes embeddings along a continuous ``quality manifold,'' improving generalization on unseen datasets. However, SCOREQ relies on heavily parameterized SSL models for feature extraction, limiting its applicability to real-time processing environments.

In this work, we bridge this gap by introducing DNSMOS-C, a novel end-to-end model that integrates a MOS-guided contrastive loss into the DNSMOS Pro framework. By applying the triplet-based SCOREQ loss on intermediate embeddings, DNSMOS-C encourages its deep latent space to better align with perceptual speech quality in a single-step optimization, maintaining the simplicity and efficiency of the framework.



The contributions in this paper can be summarized as follows. First, we present DNSMOS-C, a novel lightweight MOS prediction model that extends the DNSMOS Pro framework by integrating MOS-guided contrastive supervision directly on its intermediate latent representations. Second, we show that DNSMOS-C consistently achieves better correlation-based performance and exhibits improved robustness under domain shifts across multiple datasets. Finally, through extensive latent space analysis using PCA and clustering, we show that our model learns representations that are better organized with respect to MOS, suggesting an emergent low-dimensional quality ordering. Collectively, our experiments demonstrate that integrating MOS-guided contrastive learning enhances the performance, stability, and generalization capabilities of compact end-to-end SQA models.

\section{Method}
\label{sec:method}

\subsection{Problem formulation}

A MOS-labeled speech quality dataset consists of pairwise samples $(\mathbf{x},y)$, where $\mathbf{x}$ denotes a speech clip and $y$ its corresponding MOS. We denote the dataset as $\mathcal{D}=\{(\mathbf{x}_n\, y_n)\}_{n=1}^N$, where $N$ is the total number of speech clips in the dataset. 

The goal of an SQA model is to design a regression function $f_{\pmb{\theta}}(\mathbf{x})$ with parameters $\pmb{\theta}$ that predicts $y$. Previous works \cite{DNSMOSp, DeePMOS, DeePMOS-B} have shown that explicitly modeling a \emph{posterior distribution} $p_{\pmb{\psi}}(y|\mathbf{x})$, where $\pmb{\psi}$ are the parameters of the posterior distribution, often improves predictive performance. Using a deep neural network (DNN), we can train the regression function to predict the parameters of the posterior distribution, i.e.
\begin{eqnarray}
\pmb{\psi}(\mathbf{x}) = \mathbf{f}_{\pmb{\theta}}(\mathbf{x}). 
\end{eqnarray}
The model parameters $\pmb{\theta}$ can be learnt in a maximum-likelihood manner using the dataset $\mathcal{D}$, formulated as 
\begin{eqnarray}
    \arg \max_{\pmb{\theta}} \log \prod_{n=1}^N p_{\pmb{\psi}}(y_n|\mathbf{x}).
\label{max_likelihood_1}
\end{eqnarray}

\subsection{DNSMOS-C architecture}

\begin{figure}
    \centering
    \includegraphics[width=0.8\linewidth]{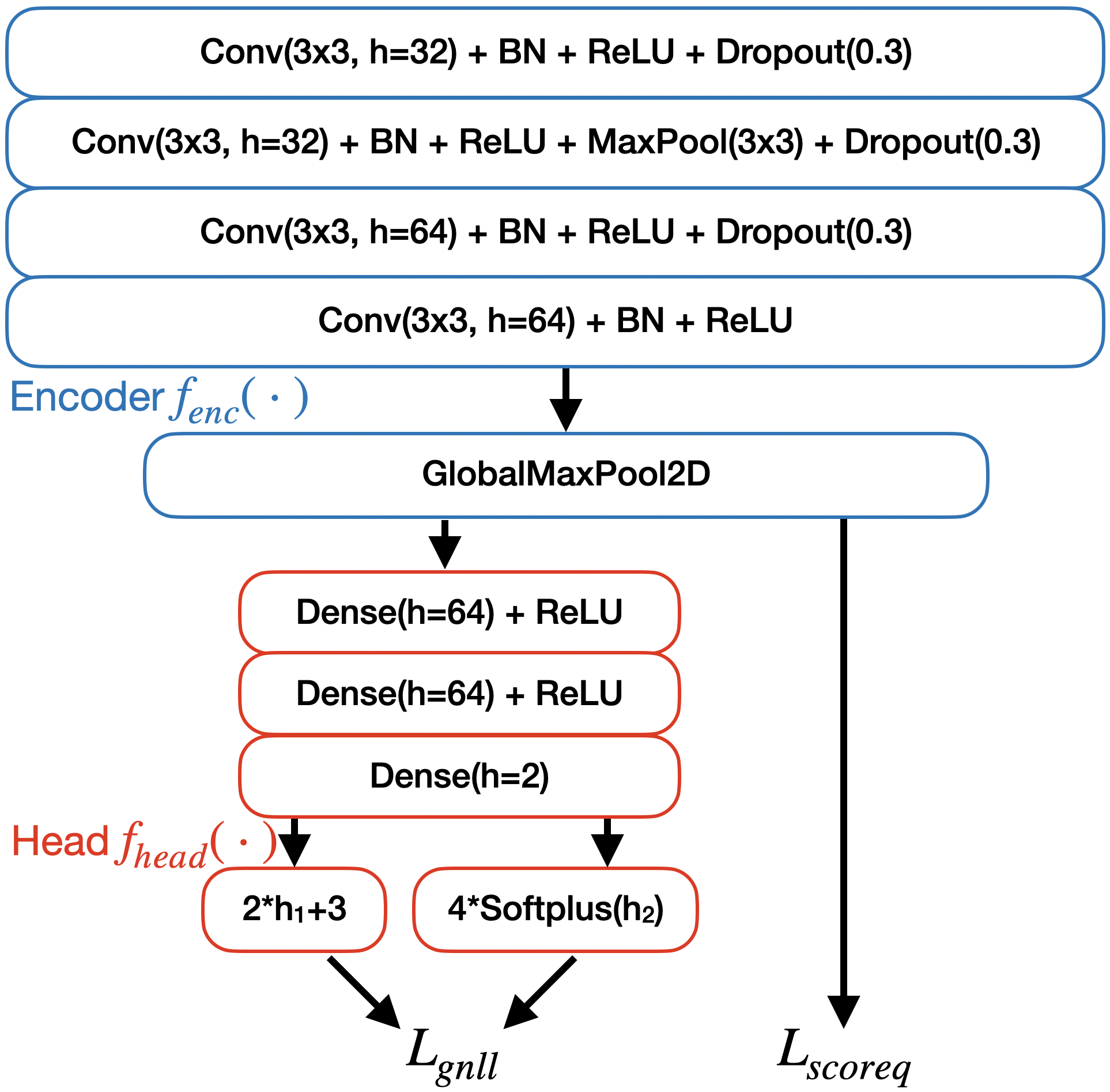}
    \caption{Architecture for DNSMOS-C}
    \label{fig:dnsmos-c}
\end{figure}

Basing our architecture on DNSMOS Pro \cite{DNSMOSp}, we assume a Gaussian posterior, i.e.,
\begin{eqnarray}
p_{\pmb{\psi}}(y|\mathbf{x}) = \mathcal{N}(y; \mu(\mathbf{x}), \sigma^2(\mathbf{x})),
\end{eqnarray}
where the network predicts both the mean $\mu(\mathbf{x})$ and variance $\sigma^2(\mathbf{x})$ for each input speech clip $\mathbf{x}$.

The training loss is then defined as the Gaussian negative log-likelihood (GNLL):
\begin{eqnarray}
L_{gnll} &=& \sum_{n=1}^N \frac{1}{2} \left[ 
\log(\sigma^2(\mathbf{x}_n)) + 
\frac{(y_n - \mu(\mathbf{x}_n))^2}{\sigma^2(\mathbf{x}_n)} 
\right].
\end{eqnarray}

The overall model architecture is illustrated in Figure~\ref{fig:dnsmos-c}, which we name \textbf{DNSMOS-C}. The model consists of an encoder module $f_{\text{enc}}$ and a head module $f_{\text{head}}$. The encoder has four convolutional layers followed by a global max-pooling layer, which maps each audio clip to a 64-dimensional embedding vector. The head module, composed of three fully-connected layers and a linear transformation, maps the embedding to the estimated MOS mean and variance.

Here, we introduce the SCOREQ loss \cite{SCOREQ}, a triplet-based contrastive regression loss, into the DNSMOS Pro framework. The original SCOREQ formulation is a multi-stage process: first, an encoder and a projection head are trained to learn a quality-aware embedding space via the contrastive loss; then, a separate regression head is fit to predict MOS scores from these embeddings.

\begin{figure}
    \centering
    \includegraphics[width=1\linewidth]{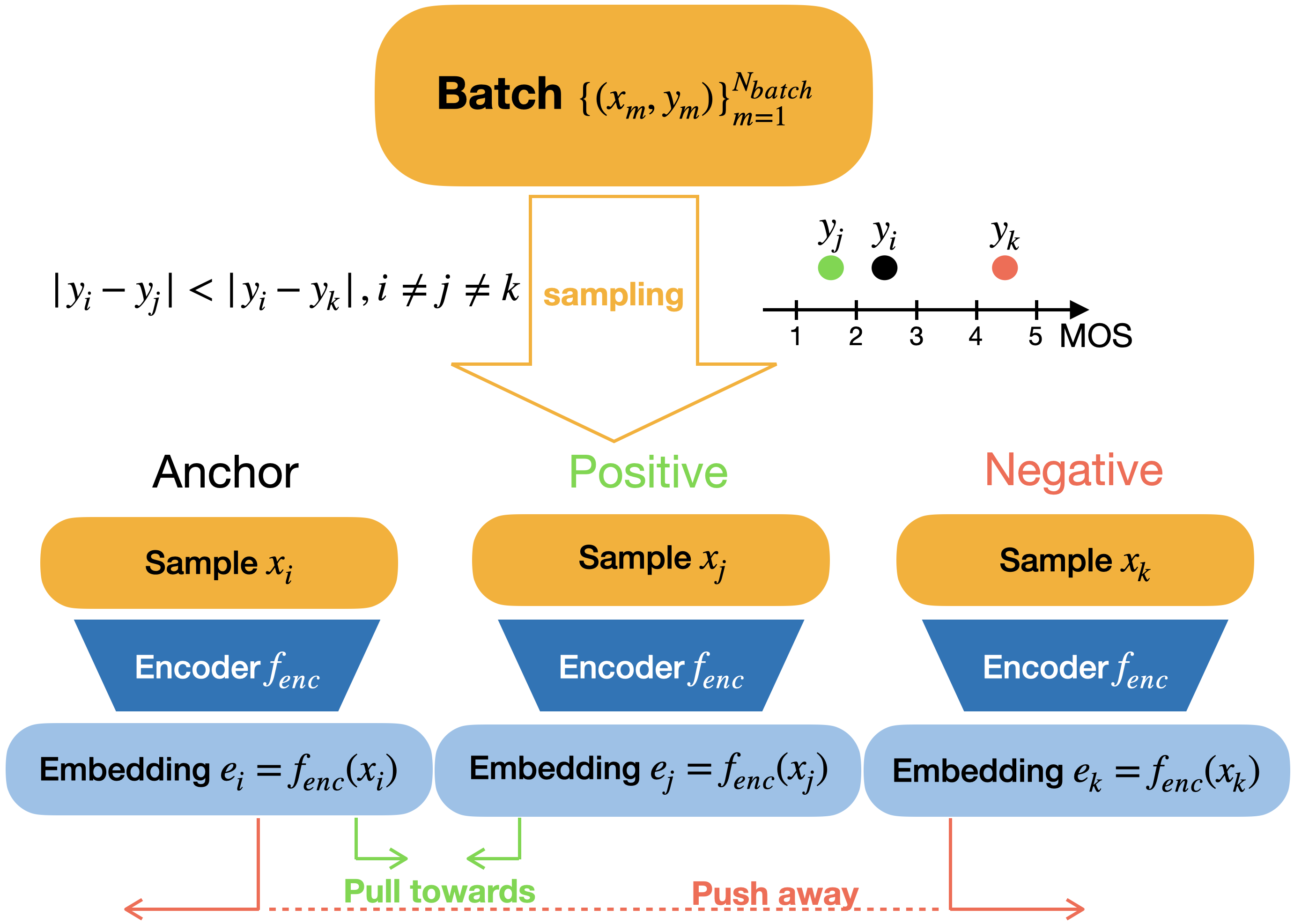}
    \caption{Illustration for SCOREQ loss in DNSMOS-C}
    \label{fig:scoreq}
\end{figure}

In contrast, our DNSMOS-C framework adopts an \emph{end-to-end} approach. Rather than depending on pre-trained SSL features or multi-stage optimization, our approach jointly guides the intermediate latent representations to be quality-aware while learning the MOS prediction task. To make SCOREQ compatible with our end-to-end setup, we directly apply the contrastive triplet loss on the embeddings produced by the encoder $f_{\text{enc}}$, and denote the embedding as $e_i$ for the $i$-th sample. This encourages the model to structure the latent space according to perceptual speech quality while predicting the MOS mean and variance. The adapted SCOREQ loss is illustrated in Figure~\ref{fig:scoreq}, and formulated as

\begin{equation}
L_{\text{scoreq}} = \sum_{\text{triplets}} \max\left(0,\; d(e_i, e_j) - d(e_i, e_k) + \delta\right),
\end{equation}
where \(d(\cdot, \cdot)\) denotes the Euclidean distance in the embedding space $e$, and \(\delta \ge 0\) is a margin enforcing separation between positive and negative pairs, in this experiment we set $\delta=0$.

\subsection{DNSMOS-C training and Inference}

Combining the GNLL loss and the adapted SCOREQ loss, the final training objective is defined as:
\begin{eqnarray}
L = L_{gnll} + \lambda \, L_{scoreq},
\end{eqnarray}
where $\lambda$ is a weighting factor that balances the contribution of the two learning objectives.

During training, the model simultaneously minimizes $L_{gnll}$ to predict the mean $\mu(\mathbf{x})$ and variance $\sigma^2(\mathbf{x})$ of the MOS distribution for each input speech clip, while $L_{scoreq}$ enforces a structured embedding space by encouraging clips with similar perceptual quality to be closer in the latent space.

At inference time, DNSMOS-C operates in a fully end-to-end manner and identical to DNSMOS Pro; given an input speech clip $\mathbf{x}$, the model predicts the Gaussian posterior parameters $\mu(\mathbf{x})$ and $\sigma^2(\mathbf{x})$. The predicted MOS is taken as the mean estimate $\hat{y} = \mu(\mathbf{x})$, and the variance estimate can serve as an uncertainty measure for downstream quality control and model confidence assessment.

\section{Experiments}
\label{sec:experiments}

\begin{table*}[t]
\centering
\begin{adjustbox}{width=\textwidth}
\begin{tabular}{cccccc}
\hline
\textbf{Dataset} & \textbf{Usage} & \textbf{Language} & \textbf{\# Samples} & \textbf{Ratings/Clip} & \textbf{Audio Source} \\
\hline
BVCC \cite{BVCC} & train/val/test & en & 4974/1066/1066 & 8 & Synthetic from TTS and VC systems \\
Tencent \cite{Tencent} & train/val/test & zh & 8000/2000/1563 & $\sim$20 & Simulated with distortions \\
NISQA\_TRAIN/VAL\_SIM \cite{NISQA} & train/val/test & en & 10000/1250/1250 & $\sim$5 & Simulated with distortions \\
NISQA\_TEST\_FOR \cite{NISQA} & test & en-au & 240 & $\sim$30 & Simulated with distortions \\
NISQA\_TEST\_P501 \cite{NISQA} & test & en-en & 240 & $\sim$28 & Simulated with distortions \\
NISQA\_TEST\_LIVETALK \cite{NISQA} & test & de & 232 & 24 & Real recordings \\
TCD-VoIP \cite{TCD-VoIP} & test & en & 384 & 24 & Simulated with distortions \\
LibriAugmented1600 \cite{cumlin2025impairments} & test & en & 1600 & - & Simulated with distortions \\
ESC50 \cite{ESC50} & test & - & 2000 & - & Real environmental sound recordings \\
\hline
\end{tabular}
\end{adjustbox}
\caption{Overview of the datasets used for training, validation, and testing.}
\label{tab:datasets}
\end{table*}

We implement DNSMOS-C \footnote{Code and checkpoints will be available at \url{https://github.com/Hope-Liang/DNSMOS-C}.} and train it on several datasets, comparing its performance against the baseline DNSMOS Pro. Three high-quality MOS-labeled datasets were used for training in various constellations. We evaluate the models on a range of test datasets to assess both generalization performance and the effect of MOS-guided contrastive learning on the latent space. All datasets used for training and evaluation are summarized in Table~\ref{tab:datasets}. The data splits for BVCC and NISQA follow the configurations used in DNSMOS Pro, while the Tencent split follows the setup in \cite{SSL-Layer-MOS}.

Data preprocessing for DNSMOS-C follows DNSMOS Pro and is identical across all datasets. All speech clips are downsampled to $16\,\text{kHz}$, and repetitively padded or cropped to $10\,\text{s}$. DNSMOS-C uses log-magnitude spectrograms as input, computed with a $20\,\text{ms}$ Hann window and $10\,\text{ms}$ hop size. The magnitudes are log-transformed and clipped to the range $[-7, 7]$.

For each of the three training datasets, namely BVCC, Tencent and NISQA\_SIM, we conduct 10 independent training runs for both DNSMOS Pro and DNSMOS-C to account for model randomness. The hyperparameter $\lambda$ in the DNSMOS-C training loss is set to $1$, determined via parameter tuning on the respective validation splits. We train both models for $500$ epochs using the Adam optimizer \cite{Adam} with a learning rate of $10^{-4}$ and moving-average parameters $\beta_1=0.9$, $\beta_2=0.999$. Model selection is performed 
based on the highest linear correlation coefficient ~\cite{LCC} on the validation set.

\subsection{Results}

We evaluate DNSMOS-C against DNSMOS Pro across test splits of the three training datasets and multiple unseen test datasets. We report the mean $\pm$ standard deviation over 10 independent runs to assess both performance and stability. Standard performance measures - mean square error (MSE), linear correlation coefficient (LCC)~\cite{LCC}, and Spearman rank correlation coefficient (SRCC)~\cite{SRCC} are used, and results are summarized in Table~\ref{tab:model_comparison} and Table~\ref{tab:model_comparison_generalization}, respectively.

\textbf{In-domain performance.}  
Table~\ref{tab:model_comparison} shows that DNSMOS-C consistently outperforms DNSMOS Pro across all training datasets in terms of correlation-based metrics. While MSE remains comparable, the consistent gains in correlation measures indicate that DNSMOS-C captures rank-order relationships between MOS ratings more effectively. The performance gain aligns with results in \cite{SCOREQ, wisnu2025improving} utilizing triplet contrastive loss.

\textbf{Generalization to unseen domains.}  
To assess cross-domain robustness, we pick models trained on NISQA\_TRAIN\_SIM and evaluate them on three unseen NISQA splits: NISQA\_TEST\_FOR, NISQA\_TEST\_P501 and NISQA\_TEST\_LIVETALK. Both models exhibit performance drops due to domain shift in language, recording environments, and speaker characteristics. However, DNSMOS-C achieves higher or on-par correlation on all the three unseen datasets. These confirm that the MOS-guided contrastive loss helps the model learn quality-aware representations that generalize better beyond training conditions, consistent with \cite{SCOREQ}.

\textbf{Training stability.}  
Across all datasets, DNSMOS-C achieves lower standard deviations than DNSMOS Pro for all performance measures, suggesting that contrastive supervision also stabilizes training. This is particularly important for small end-to-end models, which are sensitive to random initialization.

\begin{table*}[ht]
\centering
\begin{adjustbox}{width=\textwidth}
\begin{tabular}{l|ccc|ccc|ccc}
\hline
{Train Data} & \multicolumn{3}{c|}{BVCC} & \multicolumn{3}{c|}{NISQA\_TRAIN/VAL\_SIM} & \multicolumn{3}{c}{Tencent} \\
\cline{1-10}
{Model} & MSE $\downarrow$ & LCC $\uparrow$ & SRCC $\uparrow$ & MSE $\downarrow$ & LCC $\uparrow$ & SRCC $\uparrow$ & MSE $\downarrow$ & LCC $\uparrow$ & SRCC $\uparrow$ \\
\hline
DNSMOS Pro & $0.338 \pm 0.035$ & $0.791 \pm 0.016$ & $0.788 \pm 0.017$ & $\textbf{0.394} \pm 0.076$ & $0.866 \pm 0.008$ & $0.864 \pm 0.006$ & $0.282 \pm 0.046$ & $0.917 \pm 0.008$ & $0.920 \pm 0.007$ \\
DNSMOS-C & $\textbf{0.315} \pm 0.022$ & $\textbf{0.803} \pm 0.011$ & $\textbf{0.801} \pm 0.011$ & $0.424 \pm 0.068$ & $\textbf{0.868} \pm 0.005$ & $\textbf{0.868} \pm 0.004$ & $\textbf{0.259} \pm 0.032$ & $\textbf{0.921} \pm 0.005$ & $\textbf{0.925} \pm 0.003$ \\
\hline
\end{tabular}
\end{adjustbox}
\caption{Performance comparison. Metrics are reported as mean $\pm$ standard deviation, with \textbf{bold} fonts indicating better performance.}
\label{tab:model_comparison}
\end{table*}

\begin{table*}[ht]
\centering
\begin{adjustbox}{width=\textwidth}
\begin{tabular}{l|ccc|ccc|ccc}
\hline
{Test Data} & \multicolumn{3}{c|}{NISQA\_TEST\_LIVETALK} & \multicolumn{3}{c|}{NISQA\_TEST\_FOR} & \multicolumn{3}{c}{NISQA\_TEST\_P501} \\
\cline{1-10}
{Model} & MSE $\downarrow$ & LCC $\uparrow$ & SRCC $\uparrow$ & MSE $\downarrow$ & LCC $\uparrow$ & SRCC $\uparrow$ & MSE $\downarrow$ & LCC $\uparrow$ & SRCC $\uparrow$ \\
\hline
DNSMOS Pro & $\textbf{1.163} \pm 0.165$ & $\textbf{0.535} \pm 0.043$ & $0.546 \pm 0.040$ & $\textbf{0.657} \pm 0.144$ & $0.763 \pm 0.026$ & $0.758 \pm 0.030$ & $\textbf{0.935} \pm 0.208$ & $0.820 \pm 0.011$ & $0.853 \pm 0.010$ \\
DNSMOS-C   & $1.234 \pm 0.154$ & $\textbf{0.535} \pm 0.025$ & $\textbf{0.547} \pm 0.026$ & $0.686 \pm 0.135$ & $\textbf{0.787} \pm 0.027$ & $\textbf{0.784} \pm 0.026$ & $0.994 \pm 0.138$ & $\textbf{0.825} \pm 0.024$ & $\textbf{0.859} \pm 0.021$ \\
\hline
\end{tabular}
\end{adjustbox}
\caption{Generalization ability comparison on models trained on NISQA\_TRAIN/VAL\_SIM and tested on other NISQA splits.}
\label{tab:model_comparison_generalization}
\end{table*}

\subsection{Latent space analysis}

\begin{table}[h]
\centering
\begin{adjustbox}{width=\columnwidth}
\begin{tabular}{l l c c c}
\hline
\textbf{Train Data} & \textbf{Test Data} & \textbf{Metric} & \textbf{DNSMOS Pro} & \textbf{DNSMOS-C} \\
\hline
\multirow{4}{*}{BVCC} 
 & TCD-VoIP & $R$ $\uparrow$   & $0.19 \pm 0.08$ & $\textbf{0.36} \pm 0.05$ \\
 & TCD-VoIP & LCC $\uparrow$ & $0.49 \pm 0.07$ & $\textbf{0.53} \pm 0.07$ \\
 & ESC50    & Acc $\uparrow$ & $41.7 \pm 2.3$  & $\textbf{45.7} \pm 2.1$ \\
 & LA1600   & Acc $\uparrow$ & $\textbf{80.4} \pm 2.1$ & $79.0 \pm 0.8$ \\
\hline
\multirow{4}{*}{NISQA} 
 & TCD-VoIP & $R$ $\uparrow$   & $0.40 \pm 0.08$ & $\textbf{0.51} \pm 0.05$ \\
 & TCD-VoIP & LCC $\uparrow$ & $0.68 \pm 0.02$ & $\textbf{0.69} \pm 0.02$ \\
 & ESC50    & Acc $\uparrow$ & $48.0 \pm 2.3$  & $\textbf{49.7} \pm 1.3$ \\
 & LA1600   & Acc $\uparrow$ & $\textbf{85.9} \pm 0.7$ & $85.4 \pm 0.6$ \\
\hline
\multirow{4}{*}{Tencent} 
 & TCD-VoIP & $R$ $\uparrow$   & $0.45 \pm 0.04$ & $\textbf{0.49} \pm 0.05$ \\
 & TCD-VoIP & LCC $\uparrow$ & $0.52 \pm 0.04$ & $\textbf{0.53} \pm 0.07$ \\
 & ESC50    & Acc $\uparrow$ & $48.5 \pm 2.3$  & $\textbf{50.3} \pm 1.6$ \\
 & LA1600   & Acc $\uparrow$ & $\textbf{81.9} \pm 1.4$ & $79.5 \pm 1.7$ \\
\hline
\end{tabular}
\end{adjustbox}
\caption{Results for latent space analysis.}
\label{tab:model_latent_analysis}
\end{table}

\begin{figure*}[ht!]
    \centering
    \begin{subfigure}[b]{0.48\textwidth}
        \centering
        \includegraphics[width=\textwidth]{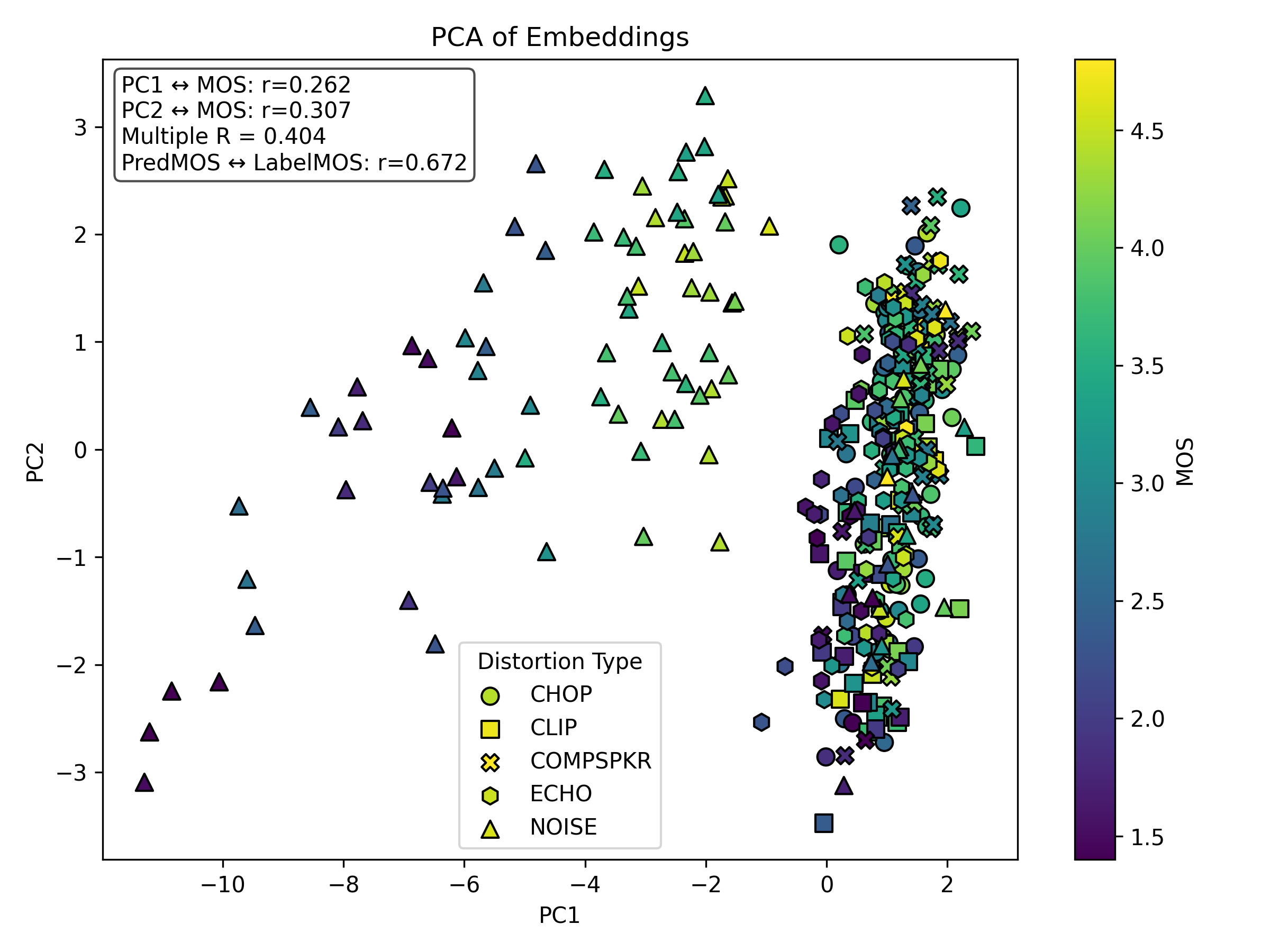}
        \caption{DNSMOS Pro}
        \label{fig:pca-sub1}
    \end{subfigure}
    \hfill
    \begin{subfigure}[b]{0.48\textwidth}
        \centering
        \includegraphics[width=\textwidth]{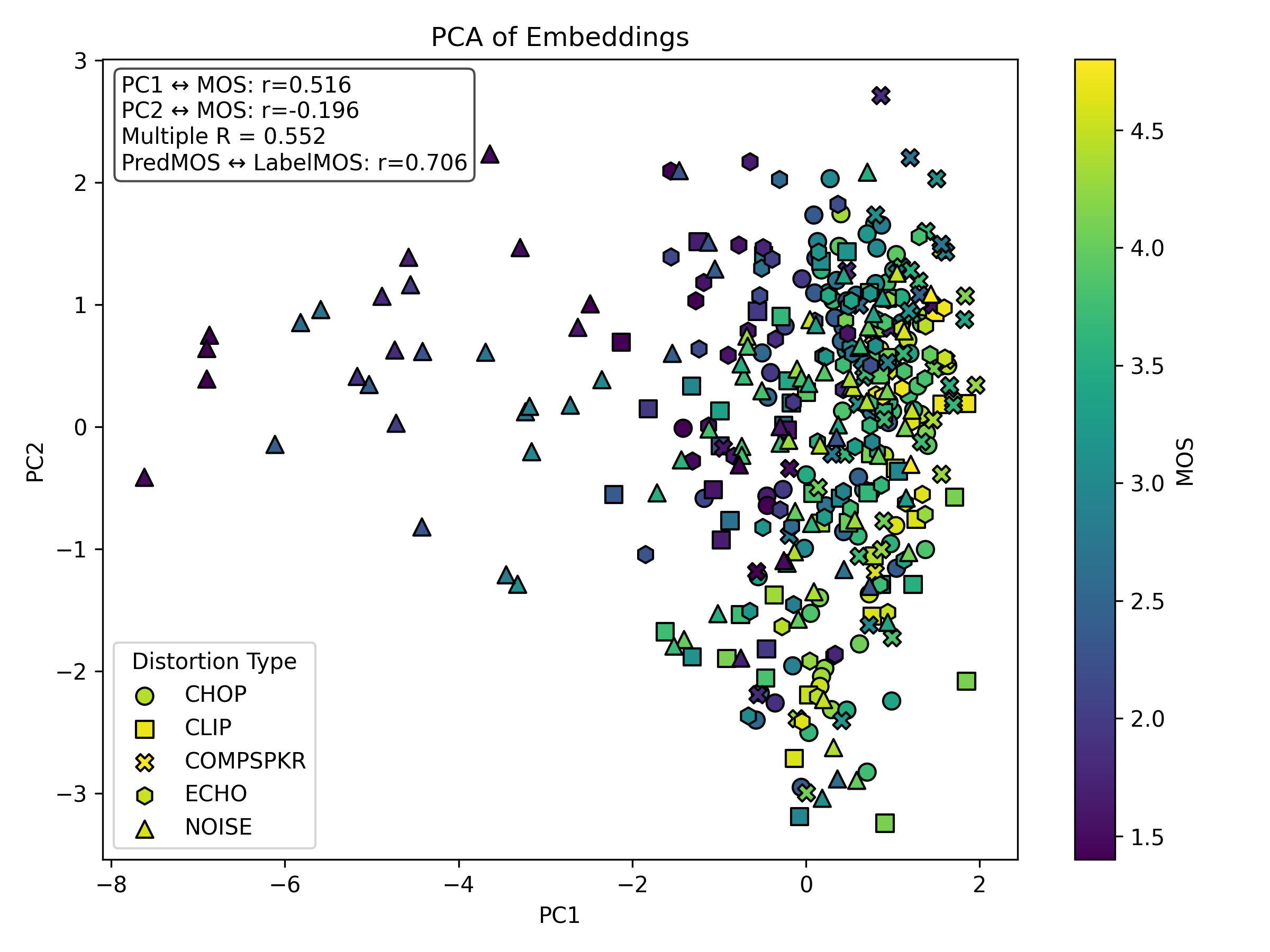}
        \caption{DNSMOS-C}
        \label{fig:pca-sub2}
    \end{subfigure}
    \caption{Latent Space Analysis on TCD-VoIP.}
    \label{fig:pca_comp}
\end{figure*}

Following the analysis in the SCOREQ paper~\cite{SCOREQ}, we investigate how the latent representations are affected by introducing the MOS-guided contrastive loss. In SCOREQ, the authors used the TCD-VoIP dataset~\cite{TCD-VoIP}, which contains five types of controlled simulated distortions with MOS annotations, to compare the latent spaces of SSL models fine-tuned with L2 loss versus SCOREQ loss. Their findings showed that the latent space fine-tuned with SCOREQ loss was reorganized primarily according to MOS scores. The two principal components (PCs) of the latent space formed a continuous ``quality manifold'', demonstrating that the contrastive loss successfully encouraged the model to learn representations that prioritize perceptual quality over specific distortion types.

In this work, we aim to verify whether a similar trend can be observed for DNSMOS-C, despite it being an end-to-end trained, significantly smaller model without extensive pretraining on large-scale data. Specifically, we conduct the following:

\begin{enumerate}
    \item \textbf{PCA-based quality correlation:} Using the unseen TCD-VoIP dataset, we apply principal component analysis (PCA) to project the latent representations into two dimensions and evaluate the multiple correlation coefficient $R$ between MOS and the two PCs jointly. We also report the LCC between the predicted MOS and the ground-truth MOS.
    \item \textbf{Impairment and noise clustering:} Following the findings of \cite{cumlin2025impairments}, which demonstrated that DNSMOS-related models naturally learn to separate distortion and noise types in their latent space, we perform a similar clustering analysis. Specifically, we use the LibriAugmented1600 (LA1600) dataset to evaluate clustering performance across 16 impairment classes, and the ESC50 dataset to evaluate clustering across 50 noise classes.
\end{enumerate}

We report the mean $\pm$ standard deviation as in previous experiments, and the overall results are summarized in Table~\ref{tab:model_latent_analysis}. For the PCA-based quality correlation, we observe that the two PCs of the latent space $f_{\mathrm{enc}}(\cdot)$ explain substantially more variance in the MOS scores after introducing the contrastive loss, as indicated by the higher multiple correlation coefficient $R$. Moreover, the LCC between predicted MOS and ground-truth MOS also improves slightly on the unseen TCD-VoIP dataset. 
Two example PCA plots from models trained on NISQA\_SIM dataset are presented in Figure~\ref{fig:pca_comp}, comparing the first two PCs between DNSMOS Pro models and DNSMOS-C models. We observe that the first principal component (PC1) becomes more strongly aligned with MOS labels shown by the color gradients in the figure, while clusters of specific distortions become less distinct. This result is consistent across the 10 random runs we tested and closely aligns with the findings reported in the SCOREQ paper~\cite{SCOREQ}. These results suggest that MOS-guided contrastive loss successfully reshapes the latent space to prioritize perceptual quality, thereby enhancing the model’s ability to generalize to unseen datasets for MOS prediction.

On the LA1600 dataset, however, we observe a slight degradation in the model’s ability to separate distortion types. This behavior is expected, as the contrastive objective explicitly optimizes the latent space for perceptual quality rather than distortion-specific discriminability, resulting in a trade-off between interpretability of distortion categories and MOS alignment. Interestingly, when evaluating clustering performance on ESC50, we observe an improvement in the separation of noise types in the latent space. We hypothesize that this is because noise characteristics are inherently correlated with MOS ratings, and thus the MOS-guided contrastive loss indirectly reinforces noise-related structure in the embeddings. In other words, while the contrastive loss reduces explicit separation among distortion categories, it enhances the organization of representations that are most predictive of perceptual quality, which in turn benefits downstream MOS prediction tasks.

\section{Conclusions}
\label{sec:conclusions}

In this work, we introduced DNSMOS-C, a novel end-to-end speech quality model that successfully integrates MOS-guided contrastive learning into the DNSMOS Pro framework. Our core contribution is a new methodology that adapts the SCOREQ triplet loss for an efficient, single-stage training pipeline, avoiding the need for pre-trained models or multi-stage optimization. This approach allows DNSMOS-C to learn a more structured, perceptually-aligned latent, which translates directly to improved performance, enhanced training stability, and better generalization to unseen datasets, all without any additional computational overhead for deployment. Future work includes testing this methodology on other end-to-end architectures to further validate its effectiveness and generalizability.

\section{Acknowledgement}
The research is supported by funding from \href{https://www.digitalfutures.kth.se/}{Digital Futures Center}, \href{https://defence-industry-space.ec.europa.eu/system/files/2023-06/REACTII-Factsheet_EDF22.pdf}{European Defence Fund REACT II} project, and partially supported by the Wallenberg AI, Autonomous Systems and
Software Program (WASP) funded by the Knut and Alice Wallenberg Foundation. The computations were enabled by resources provided by Chalmers e-Commons at Chalmers.

\section{Use of Generative AI Disclosure}

During the preparation of this manuscript, the authors used Gemini by Google to polish the language, correct grammatical errors, and improve the overall readability of the text. The authors carefully reviewed and edited all AI-generated suggestions and assume full responsibility for the final content of the paper.

\bibliographystyle{IEEEtran}
\bibliography{mybib}

\end{document}